\documentclass[10pt,letterpaper]{article}
\usepackage{opex3}
\usepackage{cite}

\begin{document}

\title{Coupling of ultrathin tapered fibers with high-Q microsphere resonators at cryogenic temperatures
and observation of phase-shift transition from undercoupling to overcoupling}

\author{Masazumi Fujiwara,$^{1,2}$ Tetsuya Noda,$^{1,2}$ Akira Tanaka,$^{1,2}$ 
\\ Kiyota Toubaru,$^{1,2}$ Hong-Quan Zhao,$^{1,2}$ and Shigeki Takeuchi$^{1,2,*}$}

\address{$^1$Research Institute for Electronic Science, Hokkaido University, Sapporo, Hokkaido 001-0021, Japan
\\
$^2$The Institute of Scientific and Industrial Research, Osaka University, Ibaraki, Osaka 567-0047, Japan}

\email{*takeuchi@es.hokudai.ac.jp} 



\begin{abstract}
We cooled ultrathin tapered fibers to cryogenic temperatures 
and controllably coupled them with high-Q microsphere resonators at a wavelength close to the optical transition of diamond nitrogen vacancy centers. 
The 310-nm-diameter tapered fibers were stably nanopositioned close to the microspheres 
with a positioning stability of approximately 10 nm over a temperature range of 7--28 K. 
A cavity-induced phase shift was observed in this temperature range, 
demonstrating a discrete transition from undercoupling to overcoupling.
\end{abstract}



\section{Introduction}
Coupling diamond nitrogen vacancy (NV) centers with nanophotonic cavities
such as microresonators and photonic crystals is currently of great importance for quantum information science
\cite{benson2011assembly, englund2010deterministic, park2006cavity, arcizet2011single, barclay2009coherent, 
santori2010nanophotonics, faraon2011resonant, faraon2012coupling, liu2011coupling, xiao2010quantum}.
Such NV-coupled cavity quantum electrodynamics (cavity QED) systems can be used to develop coherent photonic quantum devices 
including quantum phase gates \cite{kojima2003nonlinear,kojima2004efficiencies} 
and quantum memories \cite{koshino2010deterministic, wallquist2009hybrid}.
These applications exploit the narrow zero-phonon optical transitions of NV centers located at around $\lambda = 637$ nm, 
which appear only at cryogenic temperatures \cite{santori2010nanophotonics}.
To develop coherent quantum devices that employ NV centers, 
it is thus essential to couple NV optical transitions with these nanophotonic cavities in a cryogenic environment.

One of the most promising nanophotonic cavities is fiber-coupled microresonators that consist of 
high-Q silica microresonators and tapered fibers 
\cite{spillane2002ultralow, aoki2006observation, larsson2009composite, schliesser2008resolved, 
schliesser2009resolved, verhagen2012quantum, chiba2005fano, takashima2007fiber, takashima2008control, takashima2010fiber}.
Silica microresonators, such as microspheres or microtoroids, can support high-Q cavity modes ($Q > 10^8$) 
\cite{spillane2002ultralow}, 
and tapered fibers act as efficient optical interfaces between these cavity modes and waveguide modes
\cite{cai2000highly}.
To develop coherent quantum devices that use NV centers, 
the following two requirements should be realized in a cryogenic environment: 
(1) controllability of the input--output coupling efficiency between tapered fibers and microresonators 
\cite{aoki2006observation} and 
(2) well-behaved polarization characteristics adequate for measurement of a cavity-induced phase shift 
\cite{pototschnig2011controlling, tanaka2011phase}.

To satisfy the first requirement, ultrathin tapered fibers (typically 300--400 nm in diameter) have to be used 
to realize efficient coupling with microresonators at about $\lambda = 637$ nm. 
The distance between these tapered fibers and microresonators needs to be controlled with an accuracy of 10 nm. 
However, it is difficult to cool ultrathin tapered fibers to cryogenic temperatures and 
to realize such precise positioning in a cryogenic environment
due to fiber fragility and the mechanical vibrations generated by cryogenic systems \cite{park2006cavity}.
%
%
In our previous study at cryogenic temperatures \cite{takashima2010fiber}, the coupling efficiency between tapered fibers 
and microsphere cavities could not be controlled due to the large taper diameters used ($\sim$ 1.0 $\mu$m diameter) and the positioning instability.

Satisfaction of the second requirement can be demonstrated by investigating the cavity-induced phase-shift spectra.
We recently obtained phase-shift spectra of fiber-coupled microsphere resonators at room temperature by using a polarization analysis technique \cite{tanaka2011phase}. 
However, the phase-shift spectra of such fiber-coupled microresonators have not been measured at cryogenic temperatures until now. 
Phase-shift spectrum measurements at cryogenic temperatures are particularly important 
because they directly demonstrate the potential of applying these fiber-coupled microresonators to coherent quantum information devices.

Here, we demonstrate cooling of 310-nm-diameter ultrathin tapered fibers down to 7 K 
and their fully controllable coupling with high-Q microsphere resonators at about $\lambda = 637$ nm.
The tapered fibers were stably nanopositioned close to the microspheres 
with a positioning stability of less than 13 nm over at least 7 min
over a temperature range of 7--28 K. 
A cavity-induced phase shift was observed.
These observations clearly show a discrete transition 
from undercoupling to overcoupling \cite{cai2000observation, tanaka2011phase}. 
In addition, 0.8-GHz frequency tuning of the cavity resonance was realized by increasing the temperature from 7 to 28 K while preserving near-critical coupling throughout this temperature range.

The present cavity system has high-Q factor of $1.9 \times 10^7$ (potentially up to $10^9$ \cite{gorodetsky1996ultimateQ}) 
in the visible wavelength region together with the full controllability of waveguide-cavity coupling condition, 
which makes it distinct from other cryogenic fiber-coupled microcavity systems recently reported 
\cite{schliesser2008resolved, schliesser2009resolved, verhagen2012quantum, safavi2012observation}
in view that the present system can provide a testbed for quantum optics 
using stable solid-state quantum nanoemitters like diamond NV centers.
The present demonstration is thus especially important for developing coherent quantum devices 
using those solid-state quantum nanoemitters.

\section{Experiments}
\begin{figure}[t!]
\centering
	\includegraphics{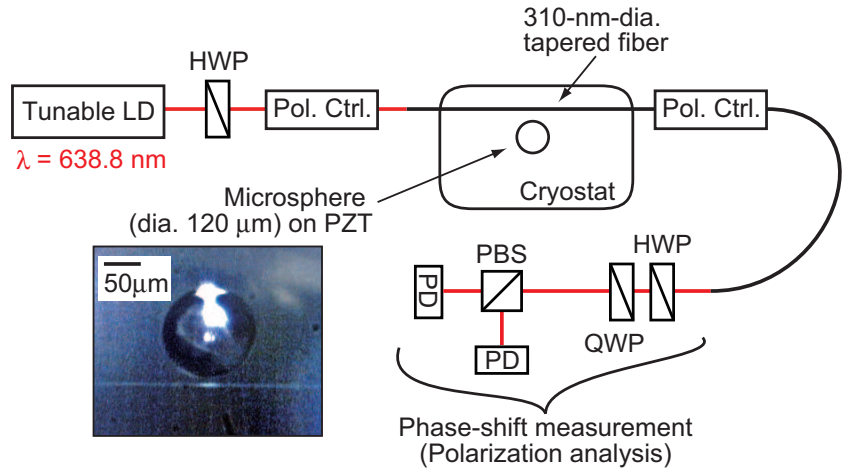}
	\caption{Schematic diagram of experimental setup. LD: laser diode; HWP: half-wave plate; Pol. Ctrl.: polarization controller;
	QWP: quarter-wave plate; PBS: polarizing beam splitter; PD: photodetector.
	The photograph shows a critically coupled fiber--microsphere system at 7 K.
	Both ends of the tapered fiber were spliced with single-mode fibers. 
	The input power to the tapered fiber was $\sim 10$ nW.
	}
	\label{fig1}  
\end{figure}

Ultrathin tapered fibers with diameters of $\sim 300$ nm were fabricated from 
conventional single-mode optical fibers (Thorlabs, 630HP) by the method described elsewhere \cite{fujiwara2011highly, fujiwara2011optical}. 
Adiabatic tapering was confirmed as the transmittance was larger than 0.9 
during fabrication.
UV adhesives were used to attach the tapered fibers to a silica substrate 
\cite{fiberloss}.
The tapered fibers used in the present experiments do not exhibit birefringence or depolarization in tapered region 
\cite{konishi2006polarization}, which ensures the polarization selectivity of cavity modes in fiber--microsphere coupling experiments.
Microspheres were fabricated by melting the tips of silica fibers (S630-HP, Thorlabs) using a CO$_2$ laser 
\cite{about-sphere-Qfactor}.

Using these ultrathin tapered fibers and microspheres, 
we performed transmittance and phase-shift measurements based on 
polarization analysis of the light in the fiber-coupled microsphere cavities.
Figure \ref{fig1} shows the experimental setup. 
The output of a tunable laser diode ($\lambda = 638.8$ nm) was coupled to a single-mode fiber, 
which was connected with a tapered fiber in the cryostat.
Polarization controllers were inserted before and after the tapered fiber to compensate the fiber birefringence. 
The output light from the fiber was then sent to the polarization analysis setup that consisted 
of a half-wave plate, a quarter-wave plate, a polarizing beam splitter, 
and two photodetectors. 
The tapered fiber used in this experiment was 310 nm in diameter. 
The microsphere was 120 $\mu$m in diameter and its stem was 20 $\mu$m in diameter.
It was mounted on a three-axis piezo stage \cite{takashima2010fiber}.
After cooling the cryostat, the microsphere was brought close to the fiber 
to couple the guided light and the cavity mode.
The polarization analysis provided four Stokes parameters from $S_0$ to $S_3$, 
which can be used to calculate the phase shift ($\Delta \phi$) due to microsphere coupling 
[$\Delta \phi = \arctan{(S_3 / S_2)}$].
The experimental details of the phase-shift measurement are described in Ref. \cite{tanaka2011phase, differtanaka}.

\section{Results and Discussions}

\begin{figure}[t!]
\centering
	\includegraphics{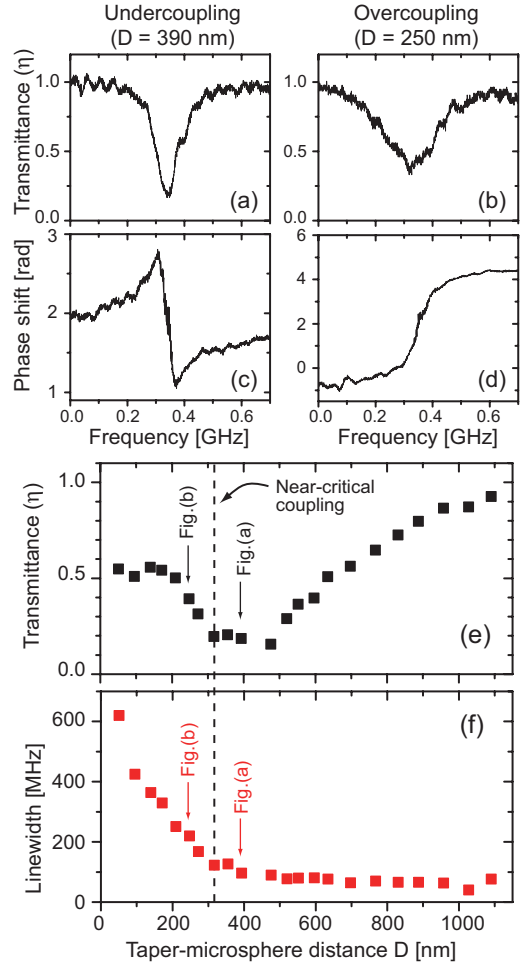}
	\caption{Transmittance and phase-shift spectra of the microsphere cavity resonance at 10 K measured 
	at taper--microsphere distances D of (a, c) 390 nm and (b, d) 250 nm, respectively. 
	Plots of the (e) transmittance minimum and 
	(f) the linewidth of the cavity resonance dip as a function of D. 
}
	\label{fig2} 
\end{figure}

Figures \ref{fig2}(a) and (b) show normalized transmittance ($\eta$) spectra of the 
cavity resonance measured at 10 K for taper--microsphere distances (D) of 390 and 250 nm, respectively.
Here, $\eta$ was normalized to the off-resonance transmittance of  the cavity mode.
The transmittance minimum and linewidth in Fig. \ref{fig2}(a) are respectively $\eta = 0.152$ and 77 MHz (Q = $6.1 \times 10^6$) 
and are respectively $\eta = 0.395$ and 220 MHz (Q = $2.1 \times 10^6$) in Fig. \ref{fig2}(b).
The higher quality factors were obtained as the microsphere went away from the tapered fiber; 
e.g. Q = $1.2 \times 10^7$ (linewidth: 40 MHz) was observed at D = 1030 nm [see Fig. \ref{fig2}(f)].

%
%
%
These two transmittance spectra exhibit different coupling conditions, 
namely undercoupling and overcoupling \cite{cai2000observation, tanaka2011phase, totsuka2006slow}.
In Fig. \ref{fig2}(a), the microsphere is located far from the fiber and 
consequently the coupling strength is smaller than the cavity loss (the undercoupling regime). 
In contrast, the microsphere is near the fiber in Fig. \ref{fig2}(b) and 
the coupling strength is greater than the cavity loss (the overcoupling regime).
Between these two regimes there is critical coupling for which the coupling strength equals the cavity loss. Critical coupling gives the highest coupling efficiency.

The striking difference between these two regimes can be clearly observed in their phase-shift spectra.
The phase shift ($\Delta \phi$) exhibits a discrete transition between these two regimes 
\cite{tanaka2011phase, totsuka2006slow}.
Figures \ref{fig2}(c) and (d) show phase-shift spectra corresponding to Figs. \ref{fig2}(a) and (b), respectively.
In the undercoupling regime, the phase shift caused by cavity resonance is less than $\pi$ and 
the phase shift does not change between the lower frequency side of the cavity resonance and the higher side.
In contrast, a phase shift of $2\pi$ is always obtained in the overcoupling regime and the phase shift changes by $2\pi$ 
as the frequency increases across the cavity resonance \cite{totsuka2006slow}.
The present observation of a discrete transition from undercoupling to overcoupling 
demonstrates the full controllability of taper--microsphere coupling in a cryogenic environment at $\lambda \sim 637$ nm.

In addition to observing these drastic changes, we have controlled the coupling condition from 
undercoupling to overcoupling by varying the taper--microsphere distance.
Figures \ref{fig2}(e) and (f) show plots of the transmittance minimum and the linewidth 
as a function of the taper--microsphere distance, respectively \cite{calibration}.
As the microsphere approaches the fiber, the transmittance minimum decreases until D = 360 nm.
Below D = 360 nm, the transmittance minimum increases until $\eta \sim 0.5$ at D = 0 nm. 
At the same time, the linewidth of the cavity resonance dip increases rapidly below D = 360 nm.
The positioning stability of the tapered fiber relative to the microsphere was estimated to be less than 13 nm over 
at least 7 minutes by calibrating the fluctuations in the transmittance minimum observed in 7 minutes 
with the distance dependence of the transmittance depicted in Fig. \ref{fig2}(e).
Such coupling controllability is crucial for developing cavity QED systems \cite{aoki2006observation}.
Thus, the present system can be effectively used for cavity QED studies using NV centers.
Note that the nonzero transmittance minimum for critical coupling in Fig. \ref{fig2}(e) originates from the phase mismatch between the waveguide and cavity modes.
This mismatch can be eliminated by optimizing the taper diameter to the microsphere coupling \cite{cai2000observation, little1999analytic}.

\begin{figure}[t!]
\centering
	\includegraphics{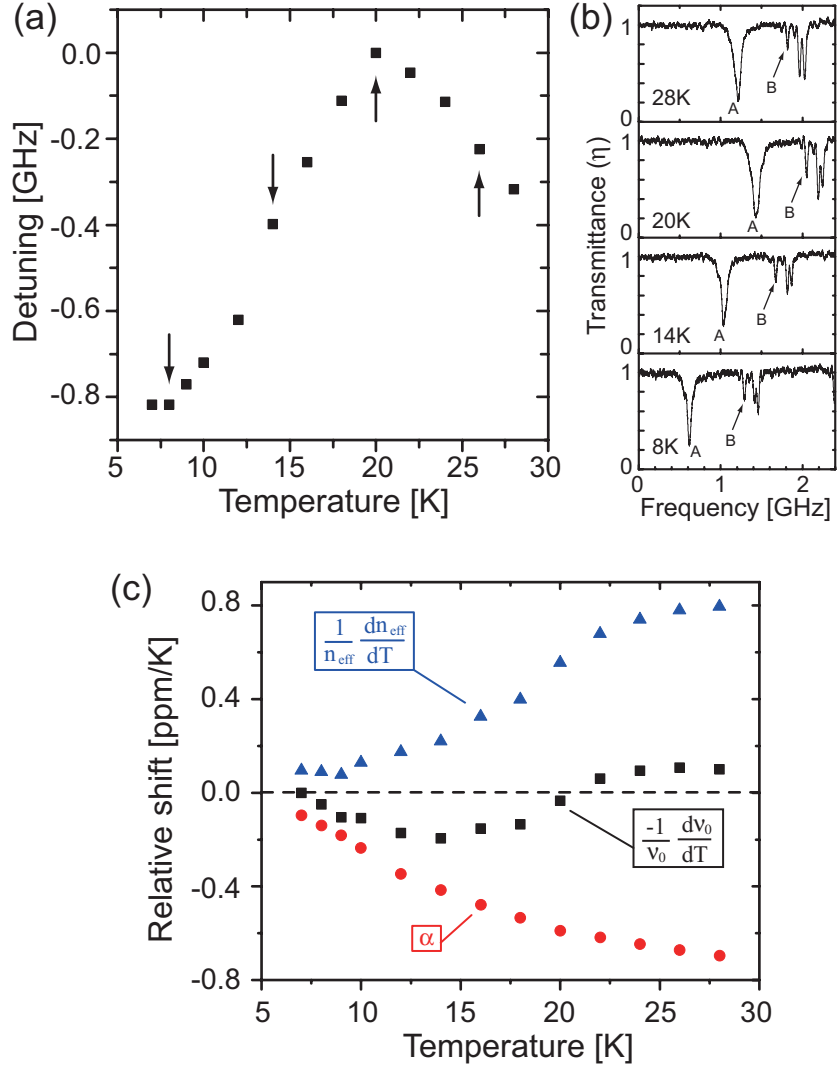}
	\caption{Cavity resonance frequency tuning by temperature change with preserving the critical coupling. 
	(a) Frequency shift of the cavity resonance as a function of temperature.
	Each data plot has an experimental error of 14 MHz that comes from the frequency fluctuation of the laser.
	(b) Corresponding transmittance spectra at 28, 20, 14, and 8 K from top to bottom [indicated by arrows in (a)].
	(c) Relative shift of $\frac{1}{\nu _0} \frac{d\nu _0}{dT}$, $\alpha$, and $\frac{1}{n_{\rm eff}} \frac{dn_{\rm eff}}{dT}$ 
	as a function of temperature from 7 to 28 K, which are indicated by black rectangle, red circle, blue triangle, respectively.
	The data of $\alpha$ is taken from Ref. \cite{whitePRL1975, whiteJPD1973}.
	In (b), cavity mode A has the Q factor of $7.3 \times 10^6$ and cavity mode B $1.9 \times 10^7$.
	}
	\label{fig3} 
\end{figure}

This coupling controllability based on stable nanopositioning of the microsphere allows their efficient coupling 
over the temperature range of 7--28 K.
Figure \ref{fig3}(a) shows the frequency shift of the cavity resonance as a function of the temperature from 7 to 28 K
and Fig. \ref{fig3}(b) shows the corresponding several transmittance spectra that 
were measured at which the microsphere was positioned to the critical coupling position.
As the temperature increases from 7 K, the central frequency of the cavity mode 
initially increases, but it starts to decrease at 20 K. 
Importantly, the transmittance minimum of the resonance dip remains constant over the entire temperature range,
as shown in Fig. \ref{fig3}(b). 
This indicates that near-critical coupling was obtained over the whole temperature range.

The frequency shift of the resonance peak showed nonlinear temperature dependence 
in contrast to the linear frequency shift at room temperature (-2.6 GHz / K at room temperature \cite{chibaJJAP2004}).
As the temperature increased from 8 K the center frequency showed positive shift up to 20 K, 
and from there it switched to negative shift.
It is known that the cavity resonance frequency shift is governed by 
thermal expansion coefficient of silica ($\alpha$) and the effective refractive index of the cavity mode ($n_{\rm eff}$) 
at cryogenic temperatures, 
which can be described as follows \cite{arcizetPRA2009, park2007regenerative},
\begin{equation}
-\frac{1}{\nu _0} \frac{d\nu _0}{dT} = \alpha + \frac{1}{n_{\rm eff}} \frac{dn_{\rm eff}}{dT}.
\label{eq1}
\end{equation}
Using the experimental values for $\frac{1}{\nu _0} \frac{d\nu _0}{dT}$ and the literature data of 
$\alpha$ \cite{whitePRL1975, whiteJPD1973}, the variation in the effective refractive index 
($\frac{1}{n_{\rm eff}} \frac{dn_{\rm eff}}{dT}$) is determined 
using this equation.
Fig. \ref{fig3}(c) shows the plot of these three components.
At room temperature, the $n_{\rm eff}$ term is known to be one order of magnitude greater than $\alpha$, 
so that $\frac{1}{\nu _0} \frac{d\nu _0}{dT}$ depends mostly on the $n_{\rm eff}$ term\cite{chibaJJAP2004, arcizetPRA2009}. 
However, the $n_{\rm eff}$ term rapidly decreases as the temperature goes down to LHe temperature regime, 
thereby competing with the thermal coefficient $\alpha$.
The resultant frequency shift of the cavity resonance peak in Fig. \ref{fig3}(c) hence shows such a nonlinear behavior (reversal of the frequency shift).

This nonlinear temperature dependence of the frequency shift in the fiber--coupled microsphere 
agrees well with the results obtained in 
Ref. \cite{park2007regenerative}, where the temperature dependence of the microsphere cavity resonance
was measured by far-field excitation and scattering detection.
This good agreement between the results of the previous study for far-field excitation and the present results for near-field excitation indicates that cavity QED experiments using NV-coupled microsphere resonators by far-field excitation, 
such as NV--photon strong coupling \cite{park2006cavity}, can be more effectively performed 
using the present near-field excitation system.

The coupling of NV centers with microcavities at cryogenic temperatures 
has been the central issue in the recent solid-state cavity QED studies. 
Until now, several works have reported the coupling of single NV centers with microcavities (cavity only)
at cryogenic temperature, demonstrating the NV-photon strong coupling \cite{park2006cavity} 
or Purcell factor of $F \sim 70$ \cite{faraon2012coupling}.
However, these microcavities still do not have efficient optical access like tapered fibers.
Forthcoming step to implement these NV-coupled cavity systems to the coherent quantum information devices is 
to provide the cavities with an efficient optical access.
Thus our present demonstration of the fiber-coupled microsphere system at cryogenic temperatures is an 
important step towards the realization of coherent quantum devices using diamond NV centers. 

The present fiber-coupled microsphere system is also distinct from 
other cryogenic fiber-coupled microcavity systems recently reported 
\cite{schliesser2008resolved, schliesser2009resolved, verhagen2012quantum, safavi2012observation} 
in view that the system is optimized for the visible wavelength range necessary for the coherent excitation of 
the stable solid-state quantum nanoemitters, including diamond NV centers or CdSe/ZnS quantum dots, and 
has the full controllability of the waveguide-cavity coupling condition with high-Q factors. 
The cavity modes that we have shown above have the Q factor of 0.6--1.2 $\times 10^7$ in the undercoupling regime 
[see Fig. \ref{fig3}(b) and cavity mode A in Fig. \ref{fig3}(b)]. 
We were also able to observe cavity modes that have Q factor of $10^7$ in many cases; 
the cavity mode B in Fig. \ref{fig3}(b) showed $Q = 1.9 \times 10^7$, for example. 
Furthermore, the microsphere-based cavity system can potentially provide Q factors up to $10^9$ \cite{gorodetsky1996ultimateQ}. 
The present system is therefore especially important for quantum optics experiments 
employing those promising solid-state nanoemitters.

\section{Summary}

We have demonstrated cooling of 310-nm-diameter ultrathin tapered fibers down to 7 K
and fully controllable coupling with high-Q microsphere resonators at $\lambda \sim 637$ nm 
in the vicinity of the zero-phonon optical transitions of diamond NV centers.
Stable nanopositioning of the 310-nm-diameter tapered fibers in close proximity to the microsphere 
with a positioning stability of less than 13 nm over at least 7 min has been realized 
over the temperature range 7--28 K. 
A cavity-induced phase shift has been successfully obtained.
The results clearly showed a discrete transition from undercoupling to overcoupling.
In addition, 0.8-GHz frequency tuning of the cavity resonance by increasing the temperature from 7 to 28 K 
has been realized while preserving near-critical coupling over this temperature range.
The present cavity systems have high-Q factor of $Q = 1.9 \times 10^7$ in the visible wavelength region 
and therefore can be used for developing coherent quantum devices using diamond NV centers.

\section*{Acknowledgments}
We thank Mr. Kamioka for his help with the fiber--microsphere coupling experiment at room temperature.
We gratefully acknowledge financial support from MEXT-KAKENHI Quantum Cybernetics (No. 21101007), JSPS-KAKENHI (Nos. 20244062, 21840003, 23244079, and 23740228), 
JST-CREST, JSPS-FIRST, MIC-SCOPE, Project for Developing Innovation Systems of MEXT, 
G-COE Program, and the Research Foundation for Opto-Science and Technology.


\begin{thebibliography}{10}
\newcommand{\enquote}[1]{``#1''}

\bibitem{benson2011assembly}
O.~Benson, \enquote{Assembly of hybrid photonic architectures from nanophotonic
  constituents,} \nat \textbf{480}, 193--199 (2011).

\bibitem{englund2010deterministic}
D.~Englund, B.~Shields, K.~Rivoire, F.~Hatami, J.~Vuckovic, H.~Park, and
  M.~Lukin, \enquote{Deterministic coupling of a single nitrogen vacancy center
  to a photonic crystal cavity,} Nano Lett. \textbf{10}, 3922--3926 (2010).

\bibitem{park2006cavity}
Y.~Park, A.~Cook, and H.~Wang, \enquote{Cavity qed with diamond nanocrystals
  and silica microspheres,} Nano Lett. \textbf{6}, 2075--2079 (2006).

\bibitem{arcizet2011single}
O.~Arcizet, V.~Jacques, A.~Siria, P.~Poncharal, P.~Vincent, and S.~Seidelin,
  \enquote{A single nitrogen-vacancy defect coupled to a nanomechanical
  oscillator,} Nature Physics \textbf{7}, 879--883 (2011).

\bibitem{barclay2009coherent}
P. E. Barclay, C. Santori, K.-M. Fu, R. G. Beausoleil, and O. Painter, 
\enquote{Coherent interference effects in a nano-assembled diamond NV center cavity-QED system,} 
\opex \textbf{17}, 8081--8097 (2009).

\bibitem{santori2010nanophotonics}
C.~Santori, P.~Barclay, K.~Fu, R.~Beausoleil, S.~Spillane, and M.~Fisch,
  \enquote{Nanophotonics for quantum optics using nitrogen-vacancy centers in
  diamond,} Nanotechnology \textbf{21}, 274008 (2010).
  
\bibitem{faraon2011resonant}
A. Faraon, P.E. Barclay, C. Santori, K.M.C. Fu, and R.G. Beausoleil, 
  \enquote{Resonant enhancement of the zero-phonon emission from a colour centre in a diamond cavity,} 
  Nature Photonics \textbf{5}, 301--305 (2011). 

\bibitem{faraon2012coupling}
A. Faraon, C. Santori, Z. Huang, V. M. Acosta, and R. G. Beausoleil, 
  \enquote{Coupling of Nitrogen-Vacancy Centers to Photonic Crystal Cavities in Monocrystalline Diamond,} 
arXiv:1202.0806v1 (2012).
  
\bibitem{liu2011coupling}
Y. C. Liu, Y. F. Xiao, B. B. Li, X. F. Jiang, Y. Li, and Q. Gong, 
  \enquote{Coupling of a single diamond nanocrystal to a whispering-gallery microcavity: 
	Photon transport benefitting from Rayleigh scattering,}
	\pra \textbf{84}, 011805 (2011).
  
\bibitem{xiao2010quantum}
Y. F. Xiao, C. L. Zou, P. Xue, L. Xiao, Y. Li, C. H. Dong, Z. F. Han, and Q. Gong, 
  \enquote{Quantum electrodynamics in a whispering-gallery microcavity coated with a polymer nanolayer,}
 \pra \textbf{81}, 053807 (2010).

\bibitem{kojima2003nonlinear}
K.~Kojima, H.~Hofmann, S.~Takeuchi, and K.~Sasaki, \enquote{Nonlinear
  interaction of two photons with a one-dimensional atom: Spatiotemporal
  quantum coherence in the emitted field,} \pra \textbf{68},
  013803 (2003).

\bibitem{kojima2004efficiencies}
K.~Kojima, H.~F. Hofmann, S.~Takeuchi, and K.~Sasaki, \enquote{Efficiencies for
  the single-mode operation of a quantum optical nonlinear shift gate,}
  \pra \textbf{70}, 013810 (2004).

\bibitem{koshino2010deterministic}
K.~Koshino, S.~Ishizaka, and Y.~Nakamura, 
\enquote{Deterministic photon-photon $\sqrt{\rm SWAP}$ gate using a $\lambda$ system,} 
\pra \textbf{82}, 010301 (2010).

\bibitem{wallquist2009hybrid}
M.~Wallquist, K.~Hammerer, P.~Rabl, M.~Lukin, and P.~Zoller, \enquote{Hybrid
  quantum devices and quantum engineering,} Physica Scripta \textbf{T137},
  014001 (2009).

\bibitem{spillane2002ultralow}
S.~Spillane, T.~Kippenberg, and K.~Vahala, \enquote{Ultralow-threshold raman
  laser using a spherical dielectric microcavity,} \nat \textbf{415},
  621--623 (2002).

\bibitem{aoki2006observation}
T.~Aoki, B.~Dayan, E.~Wilcut, W.~Bowen, A.~Parkins, T.~Kippenberg, K.~Vahala,
  and H.~Kimble, \enquote{Observation of strong coupling between one atom and a
  monolithic microresonator,} \nat \textbf{443}, 671--674 (2006).

\bibitem{larsson2009composite}
M.~Larsson, K.~Dinyari, and H.~Wang, \enquote{Composite optical microcavity of
  diamond nanopillar and silica microsphere,} Nano Lett. \textbf{9},
  1447--1450 (2009).
  
\bibitem{schliesser2008resolved}
A. Schliesser, R. Rivi{\`e}re, G. Anetsberger, O. Arcizet, and T. J. Kippenberg, 
\enquote{Resolved-sideband cooling of a micromechanical oscillator,} 
Nature Physics \textbf{4}, 415--419 (2008).

\bibitem{schliesser2009resolved}
A.~Schliesser, O.~Arcizet, R.~Rivi{\`e}re, G.~Anetsberger, and T.~Kippenberg,
  \enquote{Resolved-sideband cooling and position measurement of a
  micromechanical oscillator close to the heisenberg uncertainty limit,} Nature Physics \textbf{5}, 509--514 (2009).

\bibitem{verhagen2012quantum}
E. Verhagen, S. Del{\'e}glise, S. Weis, and A. Schliesser, and T.J. Kippenberg, 
\enquote{Quantum-coherent coupling of a mechanical oscillator to an optical cavity mode,}  
\nat \textbf{482}, 63--67 (2012). 

\bibitem{chiba2005fano}
A.~Chiba, H.~Fujiwara, J.~Hotta, S.~Takeuchi, and K.~Sasaki, \enquote{Fano
  resonance in a multimode tapered fiber coupled with a microspherical cavity,}
  \apl \textbf{86}, 261106 (2005).

\bibitem{takashima2007fiber}
H.~Takashima, H.~Fujiwara, S.~Takeuchi, K.~Sasaki, and M.~Takahashi,
  \enquote{Fiber-microsphere laser with a submicrometer sol-gel silica glass
  layer codoped with erbium, aluminum, and phosphorus,} \apl
  \textbf{90}, 101103 (2007).

\bibitem{takashima2008control}
H.~Takashima, H.~Fujiwara, S.~Takeuchi, K.~Sasaki, and M.~Takahashi,
  \enquote{Control of spontaneous emission coupling factor $\beta$ in
  fiber-coupled microsphere resonators,} \apl \textbf{92},
  071115 (2008).

\bibitem{takashima2010fiber}
H.~Takashima, T.~Asai, K.~Toubaru, M.~Fujiwara, K.~Sasaki, and S.~Takeuchi,
  \enquote{Fiber-microsphere system at cryogenic temperatures toward cavity qed
  using diamond nv centers,} \opex \textbf{18}, 15169--15173 (2010).

\bibitem{cai2000highly}
M.~Cai and K.~Vahala, \enquote{Highly efficient optical power transfer to
  whispering-gallery modes by use of a symmetrical dual-coupling
  configuration,} \ol \textbf{25}, 260--262 (2000).

\bibitem{pototschnig2011controlling}
M.~Pototschnig, Y.~Chassagneux, J.~Hwang, G.~Zumofen, A.~Renn, and
  V.~Sandoghdar, \enquote{Controlling the phase of a light beam with a single
  molecule,} \prl \textbf{107}, 63001 (2011).

\bibitem{tanaka2011phase}
A.~Tanaka, T.~Asai, K.~Toubaru, H.~Takashima, M.~Fujiwara, R.~Okamoto, and
  S.~Takeuchi, \enquote{Phase shift spectra of a fiber-microsphere system at
  the single photon level,} \opex \textbf{19}, 2278--2285 (2011). 

\bibitem{cai2000observation}
M.~Cai, O.~Painter, and K.~Vahala, \enquote{Observation of critical coupling in
  a fiber taper to a silica-microsphere whispering-gallery mode system,}
  \prl \textbf{85}, 74--77 (2000).
  
\bibitem{gorodetsky1996ultimateQ}
M. L. Gorodetsky, A. A. Savchenkov, and V. S. Ilchenko 
\enquote{Ultimate Q of optical microsphere resonators,} 
\ol \textbf{21}, 453--455 (1996).
  
\bibitem{safavi2012observation}
A.H. Safavi-Naeini, J. Chan, J. T. Hill, T. P. M. Alegre, A. Krause, and O. Painter, 
\enquote{Observation of Quantum Motion of a Nanomechanical Resonator,} 
\prl \textbf{108}, 033602 (2012).

\bibitem{fujiwara2011highly}
M.~Fujiwara, K.~Toubaru, T.~Noda, H.~Zhao, and S.~Takeuchi, \enquote{Highly
  efficient coupling of photons from nanoemitters into single-mode optical
  fibers,} Nano Lett. \textbf{11}, 4362--4365 (2011).

\bibitem{fujiwara2011optical}
M.~Fujiwara, K.~Toubaru, and S.~Takeuchi, \enquote{Optical transmittance
  degradation in tapered fibers,} \opex \textbf{19}, 8596--8601
  (2011).
  
\bibitem{fiberloss}
The overall transmittance of the tapered fibers (including the fiber coupling loss, the fiber connection loss, 
and the scattering loss at the UV adhesive) 
was 0.13 at room temperature. It decreased to 0.03 at 7 K, probably due to temperature-induced deformation of the UV adhesive. 
This reduction in transmittance does not essentially affect the present fiber--microsphere coupling experiment.

\bibitem{about-sphere-Qfactor}
Note that the microspheres fabricated from the silica fibers (S630-HP, Thorlabs) usually show Q factor of $\sim 10^7$ 
due to impurities doped in the silica fibers. 
By using high-purity and low-OH silica for the starting material,  we confirmed that 
the microspheres have Q factor of greater than $10^8$.
Note also that the cryogenic experiments do not affect the Q-factor of the microsphere as we already experimentally 
confirmed previously \cite{takashima2010fiber}.

\bibitem{konishi2006polarization}
H.~Konishi, H.~Fujiwara, S.~Takeuchi, and K.~Sasaki,
  \enquote{Polarization-discriminated spectra of a fiber-microsphere system,}
  \apl \textbf{89}, 121107 (2006).

\bibitem{totsuka2006slow}
K.~Totsuka and M.~Tomita, \enquote{Slow and fast light in a microsphere-optical
  fiber system,} \josab \textbf{23},
  2194--2199 (2006).

\bibitem{calibration}
The taper-microsphere distance was calibrated by the data of room-temperature coupling experiments
using a 330-nm-diameter tapered fiber and a 125-$\mu$m-diameter microsphere with the quality factor of $Q = 1.0 \times 10^7$.

\bibitem{differtanaka}
Equation (3) in Ref. \cite{tanaka2011phase} reads $\Delta \phi = \arctan{(S_3 / S_2)} - \arg{\rm{A_{X}}} + \arg{\rm{A_{Y}}}$, 
where $\rm{A_{X}}$ and $\rm{A_{Y}}$ are a orthogonal set of complex amplitudes of the input electric field.
In the present experiment we input diagonal polarization light, which gives $\arg{\rm{A_{X}}} = \arg{\rm{A_{Y}}} = 1/2$.

\bibitem{little1999analytic}
B.~Little, J.~Laine, and H.~Haus, \enquote{Analytic theory of coupling from
  tapered fibers and half-blocks into microsphere resonators,} J. Lightw. Technol. \textbf{17}, 704--715 (1999).
  
\bibitem{chibaJJAP2004} A. Chiba, H. Fujiwara, J. Hotta, S. Takeuchi, and K. Sasaki, 
\enquote{Resonant frequency control of a microspherical cavity by temperature adjustment,}
Jpn. J. Appl. Phys. \textbf{43}, 6138--6141 (2004).

\bibitem{arcizetPRA2009} O. Arcizet, R. Rivi\'{e}re, A. Schliesser, G. Antetsberger, and T. J. Kippenberg, 
\enquote{Cryogenic Properties of Optomechanical Silica Microcavities,} 
\pra \textbf{80}, 021803 (2009).

\bibitem{park2007regenerative}
Y.~Park and H.~Wang, \enquote{Regenerative pulsation in silica microspheres,}
  \ol \textbf{32}, 3104--3106 (2007).

\bibitem{whitePRL1975} G. K. White, \enquote{Thermal Expansion of Vitreous Silica at Low Temperatures,} 
\prl \textbf{34}, 204--205 (1975).

\bibitem{whiteJPD1973} G. K. White, \enquote{Thermal expansion of reference materials: copper, silica and silicon,}
J. Phys. D: Appl. Phys. \textbf{6}, 2070--2078 (1973).

\end{thebibliography}
\end{document}